\documentstyle[preprint,epsfig,multicol,aps,prb]{revtex}

\begin{document}
\title{Probing Spin Polarization with Andreev Reflection: A Theoretical Basis}
\author{I.I. Mazin$^1$, A.A. Golubov$^2$, and B. Nadgorny$^1$}
\address{$^1$ Naval Research Laboratory, Washington, DC, 20375\\
$^2$ University of Twente, The Netherlands }
\maketitle
\begin{abstract}
 Andreev reflection at
the interface between a ferromagnet and a superconductor has become a
foundation of a versatile new technique of measuring the spin
polarization of magnetic materials. In this paper we will briefly
outline a general theory of Andreev reflection for spin-polarized
systems and arbitrary Fermi surface in two limiting cases of ballistic
and diffusive transport.

\end{abstract}

Andreev reflection (AR) 
at the interface between a superconductor and a ferromagnet has been attracting
significant interest (e.g., Refs. \cite{sci,zutic,im99,last,gol99}) as the 
foundation of a new technique\cite{sci}
to measure the spin polarization in ferromagnets. The technique is
based on the  
idea\cite{de95} that the Andreev process is forbidden in half-metals, 
where only electrons with one spin direction are present at the
Fermi level. Correspondingly, in a ferromagnet the Andreev current is partially
suppressed, as not all of the conductance 
channels (CC) are
``open'' for AR; some channels exist in one spin direction, but not in
the other, and thus do not contribute to the Andreev current.

Building a stable and reliable
technique for probing spin polarization based on AR is not
always straightforward and requires a  quantitative
theory. Such a theory 
 should take into account the following effects:
(1) different number of CC for 
different spins, (2) finite
interface resistance, (3) band structure effects
(4) effect of an evanescent Andreev hole on quasiparticle current
in half-metallic CC,
 and (5) diffusive transport in the ferromagnet, if needed.

The existing works treat only some of these questions. The
first one was taken care of in Ref. \cite{de95}. The second
one was answered in part in the seminal paper of Blonder, Tinkham and 
Klapwijk\cite%
{BTK}, but only for ``nonmagnetic'' channels (the CC that exist in both spin
directions). The third one was dealt with in Ref.\cite{im99}, but only in the
ballistic limit. The fourth one was mentioned in Refs.\cite{zutic}, but not
investigated quantitatively. Finally, the last one was briefly touched upon
in Ref.\cite{gol99}, but left in the form that could not be directly 
applied to the experiment. Some aspects of diffusive transport and
AR were addressed in Refs. \cite{last}.

In this paper, we address all these issues and present a compendium of
formulas needed for a quantitative analysis of superconductor-ferromagnet AR.
We start with a ballistic contact, whose size is
smaller than the mean free path of electrons in the bulk. All electrons with
a positive projection of their velocity onto the current direction, 
$x$, pass through the contact. Conductance of a ballistic
contact is\cite{schep,im99}
\begin{equation}
G=\frac{e^{2}}{\hbar }\frac{1}{2}\left\langle N|v_{x}|\right\rangle A,
\label{ball}
\end{equation}%
where $A$ is the contact area, $N$ is the volume density of electronic
states at the Fermi level, $v$ is the Fermi velocity, and brackets denote
Fermi surface averaging:
\begin{equation}
\frac{1}{2}\left\langle N|v_{x}|\right\rangle =\frac{1}{(2\pi )^{3}}%
\sum_{i\sigma }\int \frac{dS_{F}}{|v_{{\bf k}i\sigma }|}v_{{\bf k}i\sigma
,x}.  \label{sharvin}
\end{equation}%
Integration and summations are over the states with $v_{{\bf k}i\sigma ,x}>0,
$ and $\Omega $ is the unit cell volume. ${\bf k},$ $i,$ and $\sigma $
denote the quasimomentum, the band index, and the electron spin,
respectively. It is instructive to look at Eq.\ref{sharvin} from the
``mesoscopic'' perspective, using as a starting point the Landauer 
formula for the
conductance of a single electron\cite{beenreview}, $G_{0}=e^{2}/h.$ The
total conductance is equal to $G_{0}$ times the number of CC, $N_{cc}$,
which is defined as the number of electrons that can pass through the
contact. If the translational symmetry in the interface plane is
not violated, then the quasimomentum in this plane, {\bf k}$%
_{\parallel },$ is conserved, and $N_{cc}$ is given by the total area of the
contact times the density of the two-dimensional quasimomenta. The latter is
$S_{x}/(2\pi )^{2},$ where $S_{x}$ is the area of the projection of the bulk
Fermi surface onto the contact plane. Thus $G=$ $\frac{e^{2}}{h}\frac{S_{x}A%
}{(2\pi )^{2}}\equiv \frac{e^{2}}{\hbar }\frac{1}{2}\left\langle
N|v_{x}|\right\rangle A$ .


Let us consider now the opposite limit, when the contact
size is much {\it larger }than the mean free path. The conductance is
then given by the the bulk conductivity, which is known from the
Bloch-Boltzmann theory:
\begin{equation}
\sigma =(e^{2}/\hbar )\left\langle Nv_{x}^{2}\right\rangle \tau ,
\label{BB}
\end{equation}%
where $\tau$ is the relaxation time. 
The Ohm's law requires that the conductance $G=\sigma A/L,$ where $L$ is
the length of the disordered region. This can be reproduced within the
``mesoscopic'' approach\cite{beenreview}, taking into account that now each
CC, that is, each separate {\bf k}$_{\parallel }$ state, has a finite
probability for an electron to get through the disordered region, $0\leq
T\leq 1,$ and
\begin{equation}
G=\frac{e^{2}}{h}\sum_{{\bf \kappa }}T_{{\bf \kappa }}=\frac{e^{2}}{h}%
\int_{\lambda }^{\infty }d\zeta P(\zeta )/\cosh ^{2}(L/\zeta ),  \label{B1}
\end{equation}%
where ${\bf \kappa \equiv \{k}_{\parallel },i,\sigma \}$. $T_{{\bf \kappa }}$
is conveniently defined in terms of the probability distribution, $P(\zeta ),
$ of the localization lengths, $\zeta $. The cutoff $\lambda $ should be of
the order of the mean free path $l;$ in fact, $\lambda =2l$\cite%
{beenreview}.
  Ohm's law requires that $G\propto 1/L,$ thus the
behavior of $P(\zeta )$ at large $\zeta $ must be $const/\zeta ^{2}.$
Normalization requires that $const=\lambda N_{cc}.$ Substituting that in Eq.%
\ref{B1}, we get
\begin{eqnarray}
G &=&\frac{e^{2}}{h}\int_{\lambda }^{\infty }\frac{\lambda N_{cc}d\zeta }{%
\zeta ^{2}\cosh ^{2}(L/\zeta ).}\approx \frac{e^{2}\lambda N_{cc}}{hL}
\label{GL} \nonumber \\
&=&\frac{e^{2}}{\hbar }\frac{A\lambda }{\Omega L}\sum_{i\sigma }\int \frac{%
dS_{F}}{|v_{{\bf \kappa }}|}v_{{\bf \kappa },x}.
\end{eqnarray}%
In the constant $\tau$ approximation, used in Eq.\ref{BB}, the
average mean free path $l=\sum_{i\sigma }\int \frac{dS_{F}}{|v_{{\bf \kappa }%
}|}v_{{\bf \kappa },x}^{2}\tau /\sum_{i\sigma }\int \frac{dS_{F}}{|v_{{\bf %
\kappa }}|}v_{{\bf \kappa },x},$ thus $\lambda _{{\bf \kappa }}=2v_{{\bf %
\kappa }x}\tau .$ Thus
$$
\left\langle G\right\rangle _{L}=\frac{e^{2}}{\hbar }\frac{A}{\Omega L}%
\sum_{i\sigma }\int \frac{2dS_{F}}{|v_{{\bf \kappa }}|}v_{{\bf \kappa }%
,x}^{2}\tau =\frac{e^{2}}{\hbar }\left\langle Nv_{x}^{2}\right\rangle \frac{A%
}{L}=\sigma \frac{A}{L}. \nonumber
$$

In the diffusive limit the conductance is determined by $\left\langle
Nv_{x}^{2}\right\rangle ,$ as it should.

The standard theory of AR (BTK)\cite{BTK}, places a specular barrier at the
interface, and assumes the ballistic regime and the free electron 
band structure
in the bulk. Let us reproduce the main results of the BTK paper 
using, instead of their derivation, the ``mesoscopic'' approach 
\cite{beenreview}%
. Probabilities of four processes must be considered: normal reflection,
AR, 
and transmission into the superconductor with or without the branch crossing%
\cite{BTK}.
The  total current can be
written as
\begin{equation}
\left\langle G\right\rangle _{NS}=\frac{e^{2}}{h}\sum_{{\bf \kappa }}T_{S}(%
{\bf \kappa )}=\frac{e^{2}}{h}\sum_{{\bf \kappa }}(1+A_{{\bf \kappa }}-B_{%
{\bf \kappa }}),  \label{TA0}
\end{equation}%
where $A$ and $B$ are the probabilities of the normal and Andreev
reflection, respectively. Beenakker showed\cite{beenreview} that the
``Andreev transparency'', the probability of an Andreev process, can be
expressed in terms of the normal transparency $T_{N}$ of the interface. For
zero bias :
\begin{equation}
T_{S}=\frac{2T_{N}^{2}({\bf \kappa )(}1+\beta ^{2})}{\beta
^{2}T_{N}^{2}+[1+r_{N}^{2}]^{2}}=\frac{%
2T_{N}^{2}(1+\beta ^{2})}{\beta ^{2}T_{N}^{2}%
+[2-T_{N}]^{2}}  \label{TA}
\end{equation}%
where $T_{N}({\bf \kappa )}$ is the normal state transparency, $r_{{\bf %
\kappa }}^{2}=1-T_{N}({\bf \kappa )}$ is the corresponding normal state
reflectance, and $\beta =V/\sqrt{|\Delta ^{2}-V^{2}|}$ is the coherence
factor. A similar formula can be derived for $V>\Delta $. For a specular
barrier, and neglecting the possible Fermi velocity mismatch at the
interface, $T_{N}({\bf \kappa })=1/[1+Z^{2}],$ where $Z$ is the BTK barrier
strength parameter \cite{BTK}. A simple algebra shows that Eq. %
\ref{TA} is equivalent to the BTK formulas.

We will now apply this approach to the diffusive AR. A diffusive 
Andreev contact
can be viewed as a contact between the normal and the superconducting 
leads, which
in addition to the interface, are separated by a diffusive region. 
The size of the region
is larger than the electronic mean free path\cite{gol99}. In the zero
temperature and zero bias limit, Eq.\ref{TA} reads:

\begin{equation}
\left\langle G\right\rangle _{NS}=\frac{e^{2}}{h}\sum_{{\bf \kappa }}T_{A}=%
\frac{e^{2}}{h}\sum_{{\bf \kappa }}\frac{2\tilde{T}_{{\bf \kappa }}^{2}}{(2-%
\tilde{T}_{{\bf \kappa }})^{2}},  \label{T-tilde}
\end{equation}%
where now the normal state transmittance for the conductance channel ${\bf %
\kappa }$ is given by the sequential conductor's formula:
\begin{equation}
{\tilde{T}}^{-1}-1 = (T_{N}^{-1}-1)+({t}^{-1}-1)
,  \label{Tt}
\end{equation}%
where $t$ is the transmittance of the diffusive region, and $T_{N}$ is the
barrier transparency. Using Eq. \ref{B1} for the distribution of $t$'s, we
find

\begin{eqnarray}
\left\langle G_{NS}\right\rangle _{L} &=&\frac{e^{2}}{h}\sum_{{\bf \kappa }}%
\frac{2}{(2/T_{N}-2+2/t_{{\bf \kappa }}-1)^{2}} \nonumber \\
&=&\frac{e^{2}}{h}\frac{\lambda N_{cc}}{L}\int_{0}^{\infty }\frac{dy}{%
[2(1-T_{N})/T_{N}+\cosh y]^{2}}.
\label{difA0}
\end{eqnarray}%
The last integral can be taken analytically and gives
\begin{equation}
\left\langle G_{NS}\right\rangle _{L}=\frac{e^{2}}{h}\frac{\lambda N_{cc}}{L}%
\frac{w\cosh w-\sinh w}{\sinh ^{3}w},
\end{equation}%
where $\cosh w=$ $2(1-T_{N})/T_{N}.$ For the clean (no-barrier) 
interface, $T_{N}=1,$
$w=i\pi /2,$ and this expression reduces to Eq.\ref{GL}, thus reproducing
the known result\cite{AVZ,beenreview} that the diffusive Andreev contact
with no interface barrier at zero bias has the same resistance in the
superconducting and in the normal states.

Is it possible then to distinguish between the spin-polarization suppression 
of the Andreev current and possible diffusive transport effect using the
experimentally measured conductance? The answer to this crucial question
is yes, as we demonstrate in Fig. 1: although it is very difficult
to discern the effect of  a finite $Z$ in a ballistic contact from
the effect of diffusive transport, it is easy to separate both
of them from the conductance suppression due to the finite
spin polarization.
 


We will now derive a full set of formulas for
arbitrary bias, temperature, and interface resistance for both ballistic and
diffusive regimes, generalizig the BTK formulas\cite{BTK}
in order to be able to use them for the half-metallic CC and in
the diffusive limit. These general formulas are summarized in Table %
\ref{2}.

We start with extending the BTK approach over the half-metallic CC, which, by
definition, correspond to the ${\bf k}_{\parallel }$ allowed in
one spin direction, but not in the other. Following BTK, we consider an
incoming plane wave
and the transmitted plane wave (with and without branch crossing)
$$
\psi _{in}=\left(
\begin{array}{c}
1 \\
0%
\end{array}%
\right) e^{ikx};\ \
\psi _{tr}=c\left(
\begin{array}{c}
u \\
v%
\end{array}%
\right) e^{ikx}+d\left(
\begin{array}{c}
v \\
u%
\end{array}%
\right) e^{-ikx},\nonumber
$$
assuming, for simplicity, the same wave vector for all the states. Here
$u$ and $v$ have the standard BTK meaning, $u^{2}=1-v^{2}=(1+\beta )/2$.
Unlike BTK, though, now the reflected state is a combination of a 
plane wave and an
evanescent wave:
\begin{equation}
\psi _{refl}=a\left(
\begin{array}{c}
0 \\
1%
\end{array}%
\right) e^{\kappa x}+b\left(
\begin{array}{c}
1 \\
0%
\end{array}%
\right) e^{-ikx}.
\end{equation}%
The total current is
$\frac{G_{HS}}{G_{0}}=\frac{4\beta \lbrack 1+(K-2Z)^{2}]}{4(1-\beta
^{2})Z(K-Z)+[1+(K-2Z)^{2}][(1+\beta )^{2}+4\beta ^{2}Z^{2}]}$ at
 $eV>\Delta $, where $K=\kappa /k,$) and 
  zero otherwise.

As $eV\rightarrow \Delta $, $G_{HS}/G_{0}\rightarrow 0$ $,$ and $%
G_{HS}/G_{0}\rightarrow G_{N}/G_{0}=1/(1+Z^{2})$ as $V\rightarrow \infty .$
We will not discuss all the aspects of the non-trivial behavior of
$G_{HS}/G_{0}$  at intermediate biases. Importantly, $G_{HS}/G_{0}$  generally
behaves non-monotonically with $V$, and may have a maximum larger than $%
G_{N}/G_{0}$ at an intermediate voltage. This maximum is due to the fact
that, although the Andreev-reflected hole does not propagate and does not
carry any current, the Andreev process itself is allowed at $eV>\Delta $ and
enhances the transparency of the barrier. This effect does not exist, though,
for $Z=0,$ nor for $K\rightarrow \infty .$
In the formulas given in Table \ref{2} we used $K\rightarrow \infty$,
to simplfy the equations,
since the actual value of $K$ matters in
a relatively narrow region of voltages above the gap.
Note that the simple renormalization of the normal current
at $eV>\Delta ,$ used in Ref.\cite{sci}, gives a rather different
result: instead of $\frac{4\beta }{\ (1+\beta )^{2}+4Z^{2}}$ it gives $\frac{%
1+\beta (1+2Z^{2})}{\ (1+\beta )(1+2Z^{2})+2Z^{4}},$ which diverges at $%
eV\rightarrow \Delta +0.$

Now we generalize the BTK formulas beyond the ballistic hypothesis. For the
nonmagnetic CC the calculation follows Eqs.\ref{TA} and \ref{Tt}. For zero
temperature and a subgap bias voltage $eV<\Delta (T)$

\begin{equation}
\left\langle G\right\rangle _{NS}=\frac{e^{2}}{h}\sum_{{\bf k}_{\parallel
},i}\frac{4\tilde{T}_{N}^{2}({\bf \kappa )}(1+\beta ^{2})}{\beta ^{2}\tilde{T%
}_{N}^{2}({\bf \kappa )}+[2-\tilde{T}_{N}({\bf \kappa )}]^{2}}
\end{equation}%
and
\begin{equation}
\tilde{T}_{N}^{-1}=T_{N}^{-1}+t^{-1}-1=Z^{2}+t^{-1},
\end{equation}%
with the distribution (\ref{B1}) for $t.$ After some  algebra we obtain
\begin{equation}
\left\langle G_{NS}\right\rangle _{L}=\frac{e^{2}}{h}\frac{\lambda N_{cc}}{L}%
\int_{0}^{\infty }\frac{(1+\beta ^{2})dy}{\beta ^{2}+(2Z^{2}+\cosh y)^{2}}.
\end{equation}%
Factor $N_{cc}$ now stands for the number of CC allowed in both spin
channels, $\lambda $ is given by the average mean free path for the channels
in question, and thus the total conductance is given by $\left\langle
Nv_{x}^{2}\right\rangle $, averaged over these channels. For $Z=0,$ this
gives
\begin{equation}
\left\langle \sigma _{NS}\right\rangle =\frac{e^{2}\tau }{\Omega }%
\left\langle Nv^{2}\right\rangle _{\downarrow \uparrow }\frac{\Delta }{V}%
\log \left| \frac{V+\Delta }{V-\Delta }\right| , \label{AVZ1}
\end{equation}%
which starts from the normal conductivity and logarithmically diverges at $%
V=\Delta$. For arbitrary $Z$ the conductance still can be cast into 
an analytical form,
namely
$$
\langle \sigma _{NS}\rangle =\frac{e^{2}\tau }{\Omega }%
\left\langle Nv^{2}\right\rangle _{\downarrow \uparrow }\frac{1+\beta }{%
2\beta }
{\rm Im}[F(-i\beta)-F(i\beta)],
$$
where
\[
F(s)=\cosh^{-1}(2Z^2+s)/ \sqrt{(2Z^2+s)^2-1}
\label{F}
\]

Similarly, for $eV>\Delta $%
\begin{eqnarray}
\left\langle G_{NS}\right\rangle _{L} &=&\frac{e^{2}}{h}\frac{\lambda N_{cc}%
}{L}\int_{0}^{\infty }\frac{2\beta dy}{\beta +(2Z^{2}+\cosh y)^{2}} \\
\left\langle \sigma _{NS}\right\rangle &=&\frac{e^{2}\tau }{\Omega }%
\left\langle Nv^{2}\right\rangle _{\downarrow \uparrow }2\beta 
F(\beta).
\end{eqnarray}%
At $Z=0$ this reduces to
\begin{equation}
\left\langle \sigma _{NS}\right\rangle =\frac{e^{2}\tau }{\Omega }%
\left\langle Nv^{2}\right\rangle _{\downarrow \uparrow }\frac{V}{\Delta }%
\log \left| \frac{V+\Delta }{V-\Delta }\right| , \label{AVZ2}
\end{equation}%
an interesting symmetry\cite{AVZnote}. At $V\gg \Delta $ we get
\begin{equation}
\left\langle \sigma _{N}\right\rangle =\frac{e^{2}\tau }{\Omega }%
\left\langle Nv^{2}\right\rangle _{\downarrow \uparrow }\frac{\cosh
^{-1}(2Z^{2}+1)}{Z\sqrt{Z^{2}+1}},
\end{equation}%
which should be used to normalize the whole conductance curve.

Finally, for for the ``half-metallic'' CC, there is no conductance at $%
eV<\Delta .$ For $eV>\Delta $,%
\begin{eqnarray*}
\left\langle G_{HS}\right\rangle _{L} &=&\frac{e^{2}}{h}\frac{\lambda N_{cc}%
}{L}\int_{0}^{\infty }\frac{2\beta dy}{\ (\beta
+1)^{2}+2(2Z^{2}-1+\cosh y)} \\
\left\langle \sigma _{HS}\right\rangle  &=&\frac{e^{2}\tau }{\Omega }%
\left\langle Nv^{2}\right\rangle _{\downarrow }\beta F[
(\beta +1)^{2}/2)-1],
\end{eqnarray*}%
where the arrow in the subscript shows that these channels are allowed only
in one spin subband.

It is again instructive to see how this expression behaves at $V\gg\Delta :$%
\begin{equation}
\left\langle \sigma _{HS}\right\rangle =\frac{e^{2}\tau }{\Omega }%
\left\langle Nv^{2}\right\rangle _{\downarrow }\frac{\cosh ^{-1}(2Z^{2}+1)}{%
2Z\sqrt{Z^{2}+1}},
\end{equation}%
which is exactly twice less than the corresponding nonmagnetic limit.

The formulas derived in this section, and summarized in Table \ref{2},
finalize our task of generalization of the BTK equations over the finite spin
polarization in both ballistic and diffusive limits. The finite temperatures
are taken into account straightforwardly in the same way as in the original
BTK paper and are not discussed here. We thank E. Demler and I. Zutic 
for useful
suggestions.


\begin{table}[tbp]
\caption{Bias dependence of the total interface current
in different regimes: BNM =
 ballistic non-magnetic\protect\cite{BTK};
BHM =  ballistic half-metallic; DNM =  diffusive non-magnetic;
DHM = diffusive half-metallic.  
$F(s)$ is defined in the text
}
\label{2}%
\begin{tabular}{ccc}
&$E<\Delta $ & $E>\Delta $\\

BNM&$\frac{2(1+\beta ^{2})}{\beta ^{2}+(1+2Z^{2})^{2}}$ &
$\frac{2\beta}{1+\beta+2Z^{2}}$ \\

BHM&0 &
$\frac{4\beta}{(1+\beta)^{2}+4Z^{2}}$ \\

DNM&$\frac{1+\beta}{2\beta}{\rm Im}\left[F(-i\beta)
-F(i\beta )
\right] $ &
$2\beta F(\beta)$ \\

DHM&0&
$\beta
F[(1+\beta)^2/2-1]
$.
\end{tabular}%
\end{table}

\begin{figure}[tbp]
\centerline{\epsfig{file=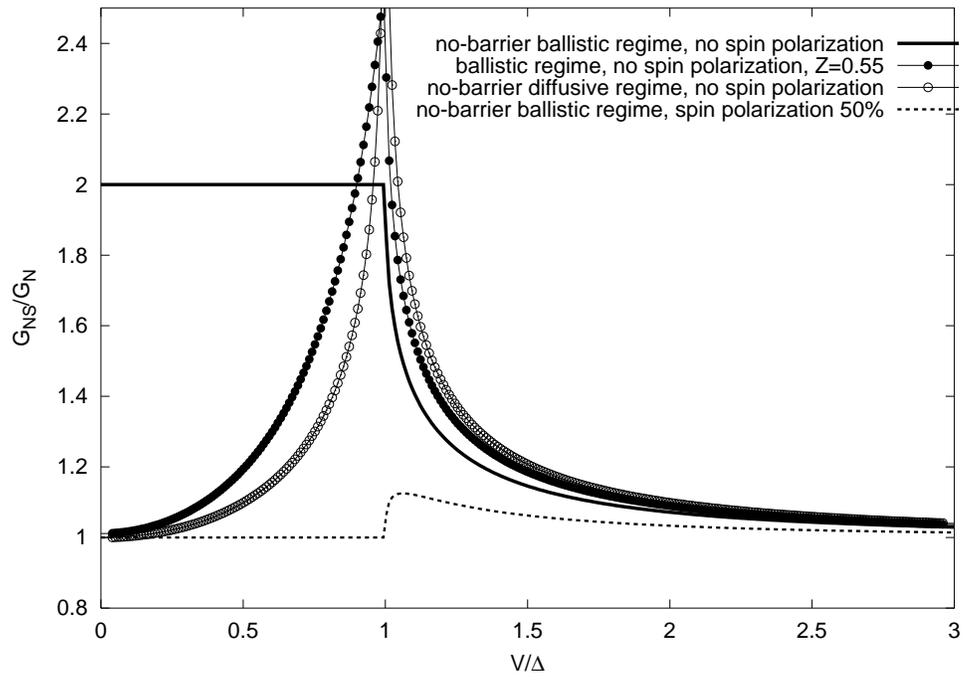,width=0.8\linewidth}}
\caption{Andreev conductance in 
in different regimes.
}
\label{modelFS}
\end{figure}

\end{document}